\documentclass[useAMS,usenatbib,usegraphicx]{mn2e}

\title[The redshift distribution of SWIFT GRBs: evidence for evolution.]{The redshift distribution of SWIFT Gamma-Ray Bursts: evidence for evolution.}
 \author[F. Daigne, E.M. Rossi
and R. Mochkovitch]{ Fr\'{e}d\'{e}ric
Daigne$^{1}$\thanks{E-mail:daigne@iap.fr}, Elena
M. Rossi$^{2}$\thanks{Chandra Fellow} and Robert Mochkovitch$^{1}$\\
$^{1}$Institut d'Astrophysique de Paris, UMR 7095 CNRS --
Universit\'{e} Pierre et Marie Curie-Paris VI, 98 bd Arago, 75014
Paris, France.\\ $^{2}$JILA, University of Colorado, Boulder, CO
80309-0440, USA}

\begin{document}

\date{Accepted **.**.**.; Received **.**.**; in original form **.**.**.}

\pagerange{\pageref{firstpage}--\pageref{lastpage}} \pubyear{2006}

\maketitle

\label{firstpage}

\begin{abstract}
We predict the redshift distribution of long Gamma-Ray Bursts (GRBs)
with Monte Carlo simulations. Our improved analysis constrains free
parameters with three kinds of observation: (i) the $\log{N}-\log{P}$ diagram of BATSE
bursts; (ii) the peak energy distribution of bright BATSE bursts; (iii)
the HETE2 fraction of X-ray rich GRBs and X-ray flashes. 
The statistical analysis of the Monte Carlo simulation results allow us to carefully study
the impact of the
uncertainties in the GRB intrinsic properties on the redshift distribution.
The comparison with SWIFT data then leads to the following conclusions.
The Amati relation should be intrinsic, if
observationally confirmed by SWIFT.
The progenitor and/or the GRB properties have to evolve to reproduce the high mean redshift of SWIFT bursts. Our results favor an evolution of the efficiency of GRB production by massive stars, that would be $\sim 6-7$ times higher at $z\sim 7$ than at $z\sim 2$.
We finally predict around
10 GRBs detected by SWIFT at redshift $z>6$ for a three year mission.
These may be sufficient to open a new observational window over the high
redshift Universe. 
\end{abstract}


\section{Introduction}
\label{sec:intro}
Gamma-ray bursts (hereafter GRBs) are powerful flashes of high-energy
photons, that travel undisturbed from cosmological distances to earth,
where they can be easily detected.  Therefore, independently of their
physical origin, GRBs can be used to probe the distant Universe \citep{lamb:00,blain:00,totani:99,wijers:98}. In
addition, there has recently been growing observational
evidence that long GRBs occur in star-forming regions
\citep{bloom:01} and that some of them are associated with peculiar
type Ic supernova explosions, i.e. with massive progenitors
\citep[e.g.][]{galama:98,stanek:03,hjorth:03,malesani:04}.  The
possible detection of GRBs at cosmological distance and their association with
massive stars motivate two questions: (i) 
how is the long GRB rate related to the star formation rate (SFR)\, ?
(ii) what is the
expected detection rate of long GRBs at high redshift\,?  On the
one hand, a better understanding of the first issue 
is a necessary step to be able to use GRBs to directly trace the star formation history in the universe.
Observations give strong indications in favor of an association of
long GRBs with massive stars.  However, the actual physical conditions
(e.g. mass, metallicity, rotation, binarity) for a star to trigger a
burst are not currently known. Thus the GRB and star formation rates may be related in a non trivial way.
On the other hand,
GRBs could be produced in the Universe in association with first stars and are thus expected up to very high redshifts \citep[e.g.][]{bromm:05}. Therefore, they are potentially important tools of investigation of the cosmic evolution;
they are probably unique direct probe at $z \ga 6.5$.  
First, the
afterglow emission may be used to study the ionization and metal
enrichment histories of the intervening intergalactic medium \citep[e.g.][]{chen:06}. 
Then, they can give insight into galaxy formation
and evolution: they allow us to detect very faint galaxies at high
redshift, that otherwise would have eluded detection \citep[e.g.][]{berger:06}.  
In turn, the detection of high redshift
galaxies can better constrain the SFR where current results are most
uncertain. Finally, they might be important to constrain the 
small-scale power
spectrum of primordial density
fluctuations \citep{mesinger:05}.\\

The standard approach to investigate such questions is to assume a GRB
comoving rate, luminosity function and spectrum
\citep[e.g.][ thereafter PM01]{porciani:01} and to constrain the free parameters of
the model by observed data, especially the GRB $\log{N}-\log{P}$
diagram \citep{schmidt:99}. We follow this approach, by the mean of Monte Carlo
simulations. Such an approach allows a realistic parametrization of
the intrinsic GRB properties, a more accurate treatment of detection
criteria for several instruments and a study of the impact of the
uncertainties in the GRB physics on the predicted GRB rate. Compared
to previous studies (\citealt{lamb:00}, PM01, \citealt{firmani:04,guetta:05,natarajan:05}), we add two more
observational constraints that allow us to better determine the model
parameters: the observed peak energy distribution of bright BATSE
bursts and the fraction of X-ray rich GRBs (hereafter XRRs) and X-ray
flashes (hereafter XRFs) in the HETE2 GRB sample. Together with the data obtained 
during SWIFT first year, this method allows us to address the two questions
that motivated this work.\\

The paper is organized as follows. In section~\ref{sec:parameters}, we
describe our assumptions on the GRB intrinsic properties (comoving
rate, luminosity and spectrum). Then, we describe our Monte Carlo
simulations in section~\ref{sec:model}. We especially detail our
detection criteria for several instruments (BATSE, HETE2 and SWIFT)
and the observations we use to constraint the free parameters of the
model. Our results are discussed in section~\ref{sec:discussion} and
section~\ref{sec:conclusion} is the conclusion.


\section{GRB intrinsic properties}
\label{sec:parameters}
Our Monte Carlo simulations are necessarily based on some assumptions
regarding the GRB intrinsic properties.  In some cases, the large
uncertainties in these properties lead us to consider several
scenarios. In this way, we can study the impact of this poor physical
understanding of the GRB phenomenon on our predictions, especially on the
GRB redshift distribution.

\subsection{Comoving rate}
\begin{figure}
  \centerline{\includegraphics[width=0.45\textwidth]{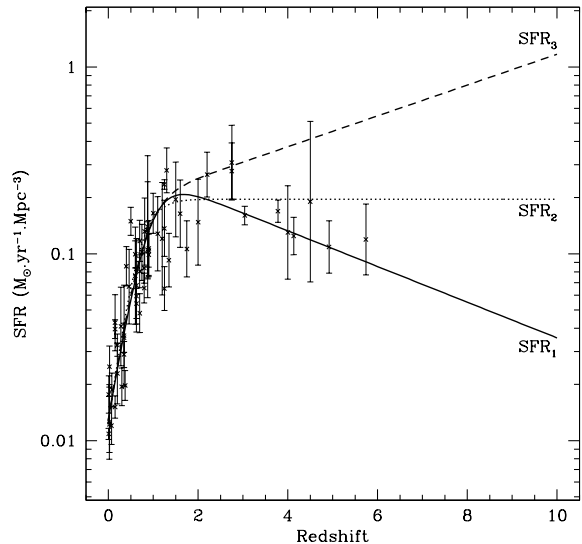}}
  \caption{The three star formation rates considered in this
  paper. Observed data are taken from \citet{hopkins:04}.}
\label{fig:sfr}
\end{figure}

Searching for high redshift galaxies has extended the number of
rest frame UV-bright systems over the interval $0\le z \la 10$
\citep[e.g.][]{giavalisco:04,bouwens:05}. These data have been used to
probe the redshift evolution of the SFR.  It seems established that
the SFR density grows by an order of magnitude from $z=0$ to $z\sim 1$
and levels off around $z\simeq 2$ (see \citet{hopkins:04} and Fig.~\ref{fig:sfr}, for a
recent collection of data).  The shape of the star formation history
is more uncertain for $z>3$.  Recent investigations of photometric
dropouts at $z\ga 6$ suggest a strong decline of more than an order of
magnitude from $z=3$ to $z \sim 10$
\citep{bouwens:03,bouwens:05,fontana:03,stanway:04}.  However, the
assessment of this result is subject to the poorly constrained dust
obscuration at such high redshifts.\\

Since long GRBs are very likely related to the
death of massive stars, the most simple assumption is to adopt a GRB
rate proportional to the SFR. Following PM01, we take
\begin{equation}
\mathcal{R}_\mathrm{GRB} = k\times \mathcal{R}_\mathrm{SN}\ ,
\end{equation}
where $\mathcal{R}_\mathrm{GRB}$ is the GRB comoving rate and
$\mathcal{R}_\mathrm{SN}$ is the type II supernova comoving rate.
We assume the same initial mass function (IMF) as in PM01. As this IMF is constant with time, the supernova rate is proportional to the SFR\,:
\begin{equation}
\mathcal{R}_{SN}=0.0122\ \mathrm{M_{\odot}^{-1}} \times \mathrm{SFR}\ ,
\end{equation}
the lifetime of massive stars being neglected.
We first considered for the SFR a
simple fit of observed data up to $z\sim 6$ \citep{hopkins:04}. This
case is thereafter called  $\mathrm{SFR}_{1}$. As the true evolution of
the SFR above $z\sim 2-3$ is still uncertain, we have also considered
two alternative rates\,: $\mathrm{SFR}_{2}$, where the SFR is constant
above $z\sim 2$ and $\mathrm{SFR}_{3}$, where the SFR still increases
above $z \sim 2$. In this last case, we arbitrarily cut the GRB rate
at $z_\mathrm{max}=20$. These three SFRs are plotted in
Fig.~\ref{fig:sfr}. They are in fact very close to those adopted
by PM01. \citet{lamb:00} considered two cases, one
being close to SFR$_{1}$ and one being close to SFR$_{2}$ with an
additional very intense component at $z\ga 6-7$. The preferred SFR in
the results by \citet{natarajan:05} is close to our SFR$_{3}$ up to
$z\sim 6-7$. Finally \citet{firmani:04} adopted a rather different
approach.  The SFR in their study is a three free parameters function.
It is adjusted together with the luminosity function under the
joined constraint of the BATSE $\log{N}-\log{P}$ diagram and the
``observed'' GRB redshift distribution provided by the
luminosity-variability relation \citep{fenimore:00}.

The ratio of GRB explosions to type II supernovae, $k$, is a free
parameter of the model. Here $\mathcal{R}_\mathrm{GRB}$ has to be
understood as the comoving rate of GRBs pointing towards us. The true
GRB rate in the Universe (as well as the true GRB / SN ratio) is
obtained by multiplying this rate by a correcting factor for the
beaming, which is of the order of $\left\langle
\left(\Omega/4\pi\right)^{-1}\right\rangle \sim 500-1000$
\citep{frail:01}.


\begin{figure*}
\begin{center}
\begin{tabular}{cc}
\textbf{$\mathbf{\log{N}-\log{P}}$ only} & \textbf{$\mathbf{E_\mathrm{p}}$ only}\\
\includegraphics[width=0.49\textwidth]{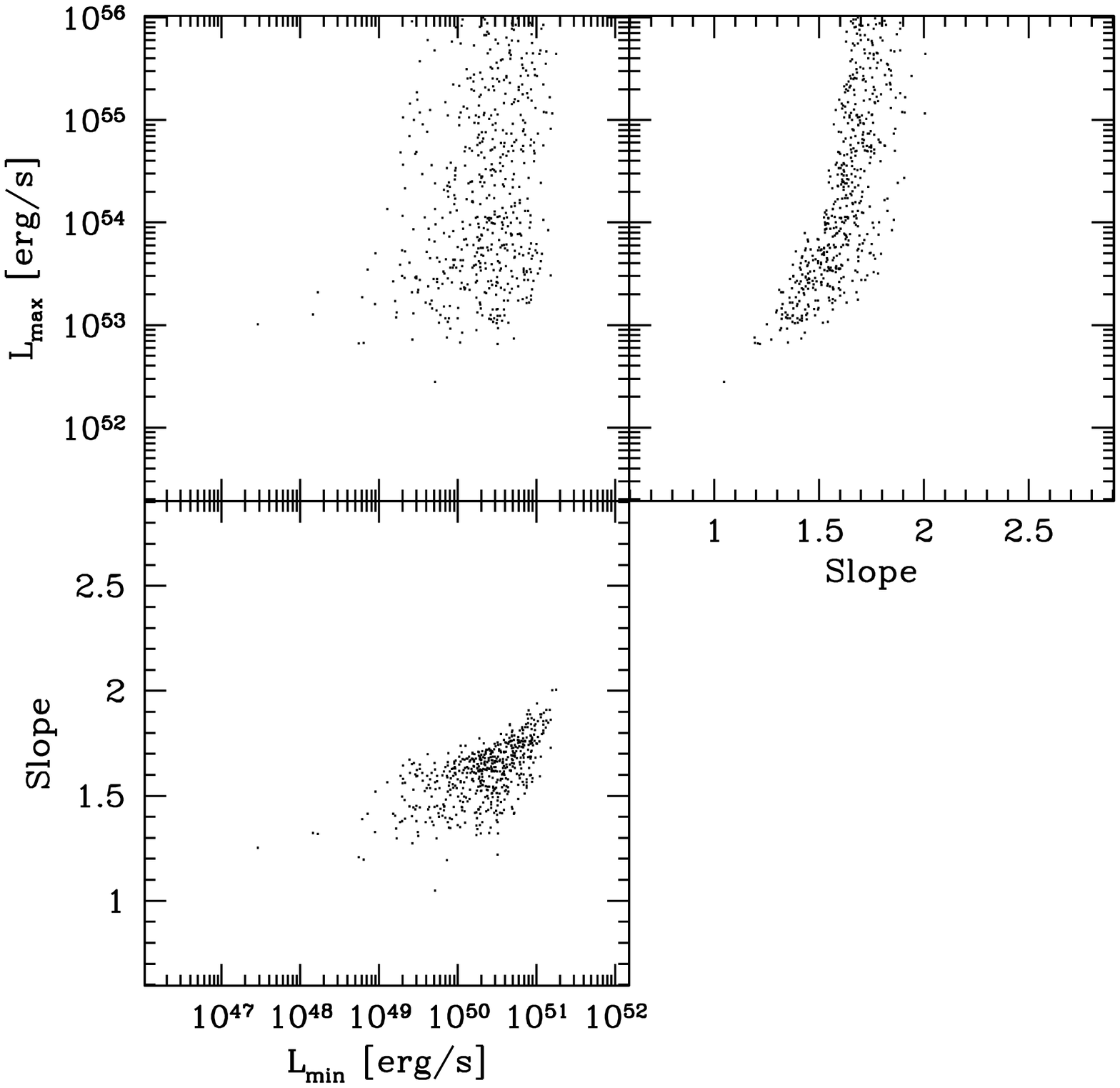}  &
\includegraphics[width=0.49\textwidth]{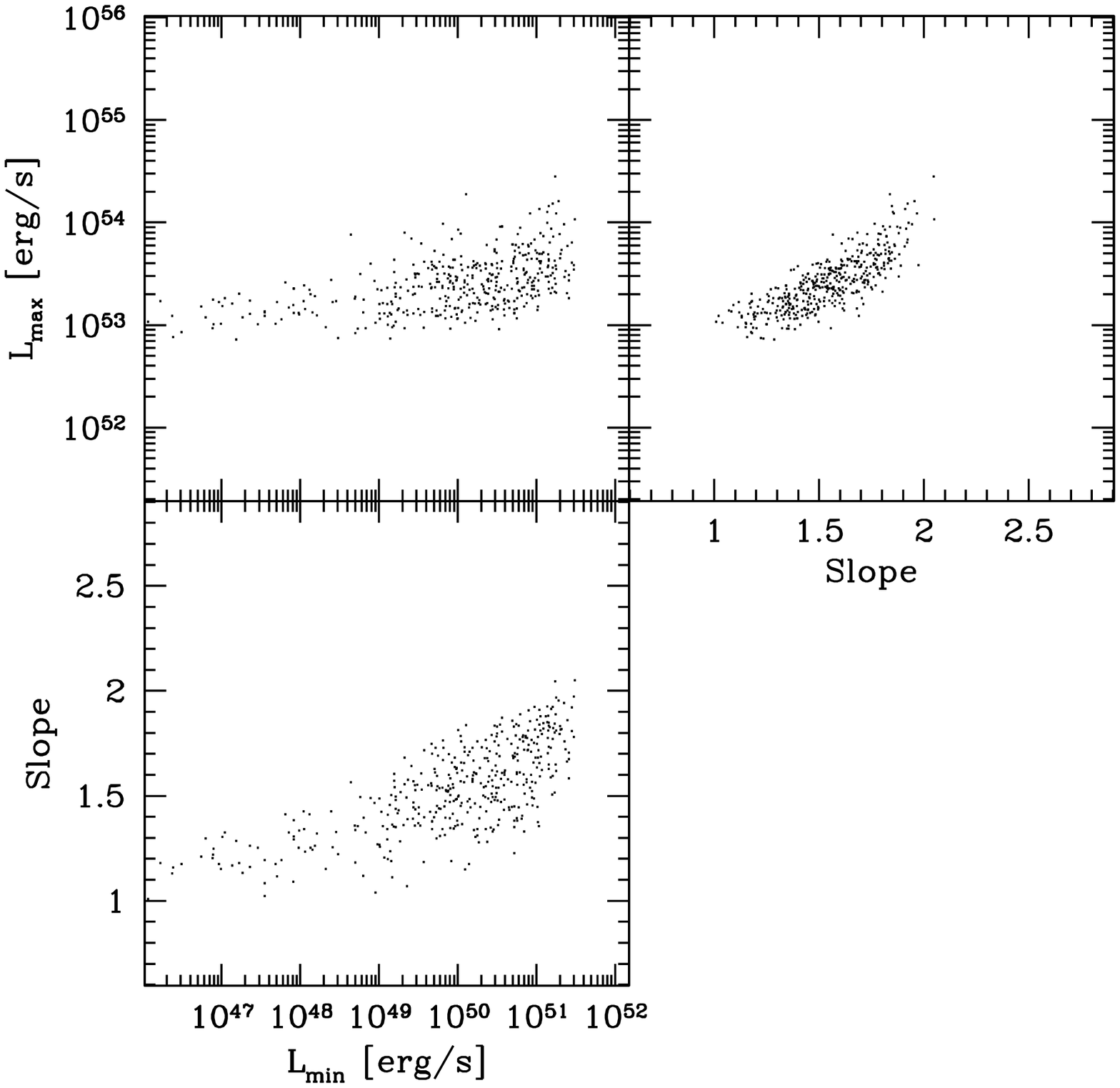} \\
\textbf{XRF+XRR fraction only} & \textbf{All constraints}\\
\includegraphics[width=0.49\textwidth]{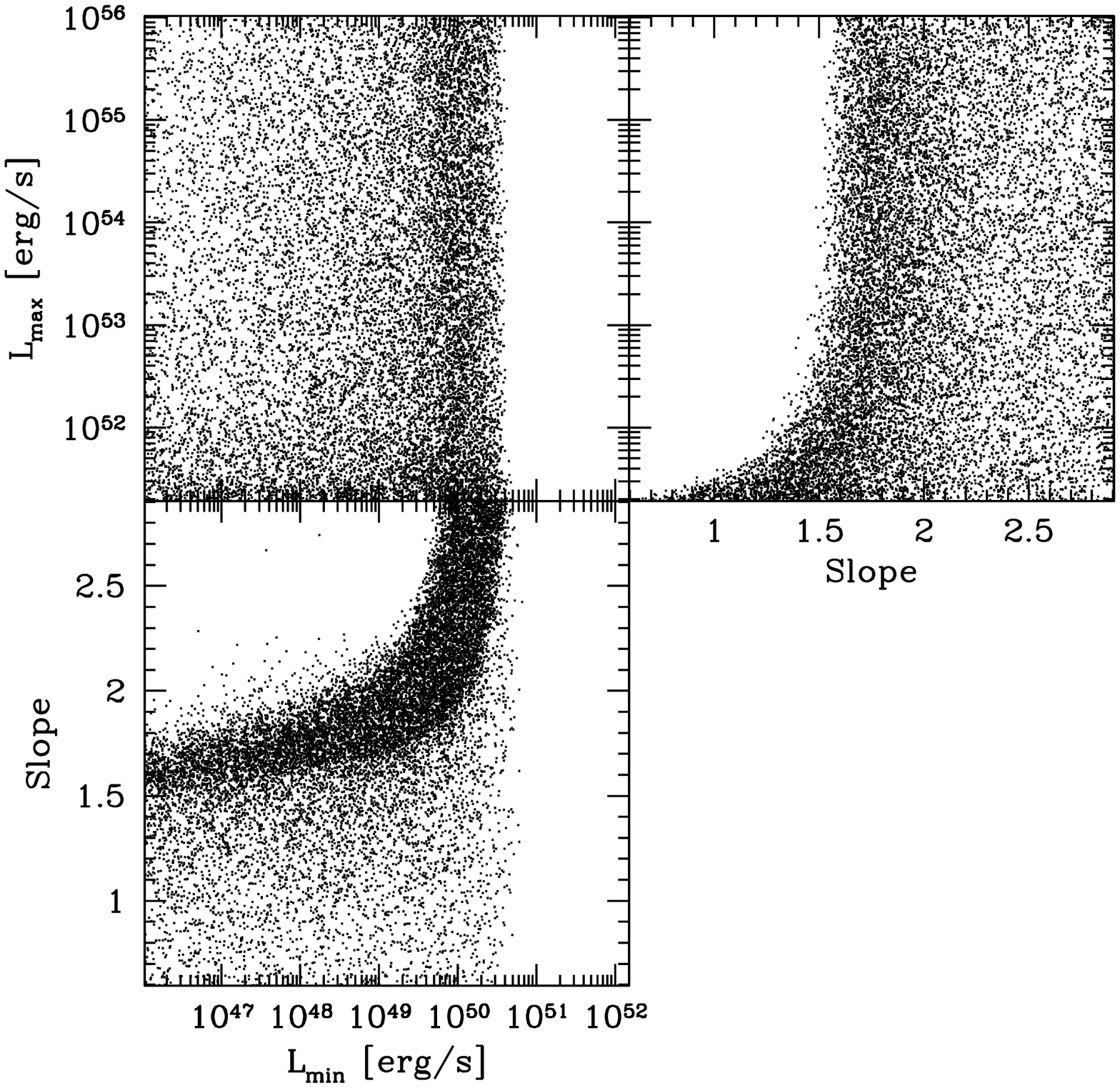}  &
\includegraphics[width=0.49\textwidth]{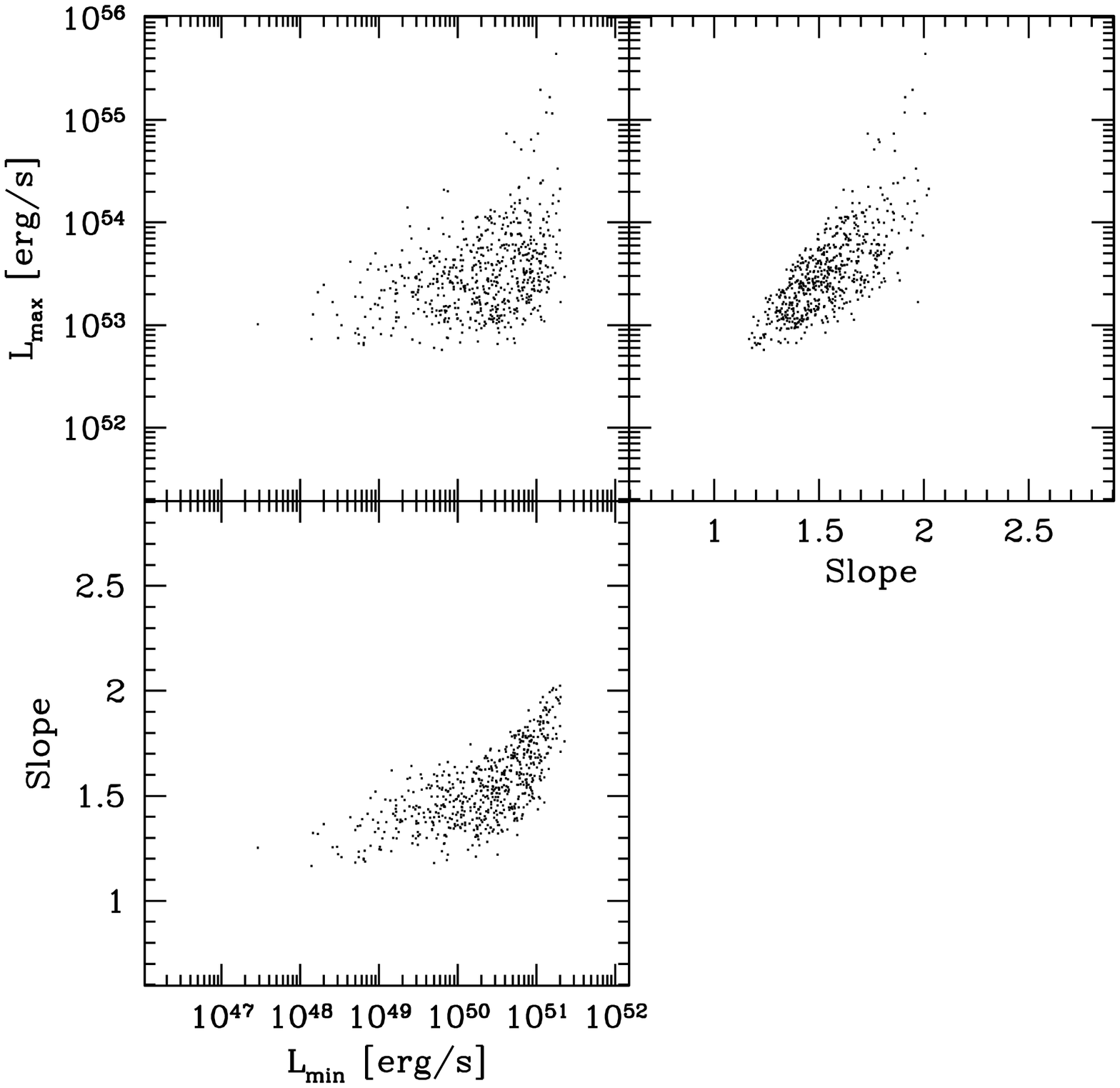} \\
\end{tabular}
\end{center}
\caption{\textbf{Parameter space (SFR$_{3}$, Amati-like relation)~:}
we plot the location in the parameter space of the best models (see section~\ref{sec:fit}). \textit{Top-left:} models that fit
the BATSE $\log{N}-\log{P}$ diagram \citep{stern:00,stern:02};
\textit{Top-right:} models that fit the peak energy distribution of
bright BATSE bursts \citep{preece:00}; \textit{Bottom-left:} models
that fit the fraction of XRFs and XRRs observed by HETE2
\citep{sakamoto:05}; \textit{Bottom-right:} models that fit the three
constraints.}
\label{fig:param}
\end{figure*}

\subsection{Luminosity function}

The GRB luminosity function is currently quite uncertain. We assume a
 power-law distribution, which does not vary with redshift:
\begin{equation}
p(L) \propto L^{-\delta}\ \mathrm{for}\ L_\mathrm{min} \le L \le
L_\mathrm{max}\ .
\end{equation}
Notice that $p(L)$ is the distribution of the isotropic equivalent luminosity of GRBs that would be obtained by a perfect (no threshold) detector on Earth. This is of course different from the true distribution of $L$ for all GRBs in the Universe, which could be obtained from $p(L)$ after correction for viewing selection effects. With $p(L)$ there are
three more model parameters: the slope $\delta$
and the minimum and maximum luminosities $L_\mathrm{min}$ and
$L_\mathrm{max}$. Previous studies also used such a power-law luminosity
function \citep{lamb:00} or a Schechter luminosity function (PM01,
\citealt{natarajan:05}). This latter is very close to a single
power-law extending to $L_\mathrm{max}=+\infty$ with an additional
low-luminosity tail below $L_\mathrm{min}$ that does not contribute much
due to an exponential cutoff. \citet{firmani:04} also used a power-law
distribution, but including the additional possibility that the
typical GRB luminosity scales with $(1+z)^{\nu}$ with $\nu\sim
0.8-1.2$. We also briefly tested such evolutionary effect (see
section~\ref{sec:redshift}).

\subsection{Spectrum}
The intrinsic photon spectrum is assumed to be a broken power-law with
a break at energy $E_\mathrm{p}$ \citep{band:93} and a low
(resp. high) energy slope $-\alpha$ (resp. $-\beta$). We checked that
a more realistic spectrum shape, the so-called ``GRB''  function
\citep{band:93}, has only a small impact on our results and does not
affect our conclusions.
The probability distributions of the
low-energy and high-energy slopes follows the observed distribution of
$\alpha$ and $\beta$ in a set of long bright GRBs
\citep{preece:00}. We have checked a posteriori that the simulated
distributions of the ``observed'' low- and high-energy slopes in long
bright GRBs were very close to the intrinsic distributions.

As the observed peak energy distribution of these same bright burst is
close to log-normal \citep{preece:00}, the most simple assumption is
to adopt an intrinsic log-normal distribution for $E_\mathrm{p}$, with
a dispersion $\sigma$ and a mean value $E_\mathrm{p,0}$. This is our
first scenario (hereafter ``log-normal $E_\mathrm{p}$
distribution''). We assume $\sigma=0.3\ \mathrm{dex}$ (we checked that
our results do not depend too much on this assumption) and we keep
$E_\mathrm{p,0}$ as a free parameter. On the other hand, there are
some evidences that the intrinsic peak energy could be correlated with
the intrinsic luminosity. Therefore, we have considered a second
scenario (hereafter ``Amati-like relation''), where
\begin{equation}
E_\mathrm{p} = 380\ \mathrm{keV}\ \left(\frac{L}{1.6\times 10^{52}\
\mathrm{erg~s^{-1}}}\right)^{0.43}\ ,
\end{equation}
with a normal dispersion $\sigma=0.2\ \mathrm{dex}$, in agreement with
the observed relation (\citealt{yonetoku:04,ghirlanda:05}, see however \citealt{nakar:05,band:05} who tested this relation against BATSE data and concluded that selection effects were dominant).
In this second scenario, no free
parameter is introduced for the spectrum.

Notice that the treatment of the spectrum in this work is improved
compared to previous studies: \citet{lamb:00} adopted a single
power-law spectrum of slope $\alpha=-1$ which of course is a very poor
approximation for high redshift GRBs. Other studies (PM01, \citealt{firmani:04,natarajan:05}) adopt a standard GRB
spectrum with fixed slopes $\alpha=-1$ and $\beta=-2.25$ and a
constant break energy $E_\mathrm{b}$ in the GRB rest frame which is a priori fixed
to 511 keV ($E_\mathrm{p}=E_\mathrm{b}\times
(2+\alpha)/(\alpha-\beta)\sim 410\ \mathrm{keV}$ in the GRB rest
frame), without any a posteriori check that the simulated observed
$E_\mathrm{p}$ distribution is in agreement with BATSE
data. \citet{firmani:04} also consider a case including an intrinsic
Amati-like relation of slope 0.5, but without any dispersion.


\begin{table}
\begin{minipage}{0.5\textwidth}
\caption{\textbf{Best models: parameters.}}
\begin{tabular}{cllll}
\hline
\textbf{SFR} & $\log{L_\mathrm{min}}$ & $\log{L_\mathrm{max}}$ & $\delta$ & $\log{k}$ \\ 
\hline
\multicolumn{5}{c}{\textbf{Amati-like relation $E_\mathrm{p}\propto L^{0.43}$}}\\ 
& & & &  \\ 
\textbf{1} & $49.9\pm 0.5$ & $53.7\pm 0.4$ & $1.70\pm 0.08$ & $-5.4\pm 0.3$ \\
\textbf{2} & $50.0\pm 0.5$ & $53.7\pm 0.5$ & $1.68\pm 0.10$ & $-5.5\pm 0.3$ \\
\textbf{3} & $50.3\pm 0.7$ & $53.5\pm 0.4$ & $1.54\pm 0.18$ & $-6.0\pm 0.2$ \\
\hline
\multicolumn{5}{c}{\textbf{log-normal distribution peak energy distribution}}\\ 
& & & &  \\
\textbf{1} & $50.2\pm 0.9$ & $53.6\pm 0.8$ & $1.62\pm 0.18$ & $-5.6\pm 0.3$ \\
\textbf{2} & $50.2\pm 1.1$ & $53.6\pm 0.9$ & $1.62\pm 0.27$ & $-5.7\pm 0.3$ \\
\textbf{3} & $50.5\pm 1.3$ & $53.7\pm 0.9$ & $1.52\pm 0.48$ & $-6.2\pm 0.2$ \\
\hline
\end{tabular}\\
\begin{tabular}{cl}
\hline
\textbf{SFR} &  $\log{E_\mathrm{p,0}}$\\ 
\hline
\multicolumn{2}{c}{\textbf{log-normal distribution peak energy distribution}}\\ 
& \\
\textbf{1} & $2.74\pm 0.08$ \\
\textbf{2} & $2.73\pm 0.08$ \\
\textbf{3} & $2.79\pm 0.08$ \\
\hline
\end{tabular}
\end{minipage}
\label{tab:bestfit}
\end{table}

\section{Monte Carlo simulations}
\label{sec:model}
In our Monte Carlo simulations, each generated GRB is given a
redshift, a luminosity and a spectrum, according to the specific
intrinsic distributions that have been described above. Then
observed properties (observed peak energy, peak flux) can be computed
and compared to the sensitivity of different instruments. In this
study, we focus on BATSE, HETE2 and SWIFT. Our detection criteria are
specified in section~\ref{sec:instruments}. This allows us to derive
the expected distributions of observed redshifts, peak fluxes and peak
energies for these instruments, for comparison with real data. It also
allows to reconstruct the source frame properties
(e.g. luminosity, peak energy) of detected bursts. 
In a second step, the free parameters of the model are
adjusted to fit several observational constraints. This is explained
in section~\ref{sec:obs}.

\subsection{Detection criteria}
\label{sec:instruments}
We consider the detection by three instruments: BATSE, HETE2 and
SWIFT. For BATSE, we apply the detection efficiency as a function of
the peak flux, $\epsilon(P)$, given by \citet{kommers:00}, so that we can
compare our results with the $\log{N}-\log{P}$ diagram published by
these authors. We also compare our simulated bursts with the
$\log{N}-\log{P}$ diagram published by \citet{stern:00,stern:02}. 
Their sample goes farther towards low flux bursts and puts therefore
more constraints on the GRB rate at high redshift. Their
published diagram is already corrected for detection efficiency. In
addition we apply a threshold $P_{50-300\ \mathrm{keV}} > 5\
\mathrm{ph~cm^{-2}~s^{-1}}$ defining a sub-sample of ``bright BATSE
bursts'', to compare our simulated GRBs with the results of the BATSE
spectroscopic catalog \citep{preece:00}.  
In our simulations, 
this threshold typically
corresponds to 5 \% to 10 \% 
of BATSE bursts,
in agreement with
observations.
For HETE2, we adopt the same threshold of
$1\ \mathrm{ph~cm^{-2}~s^{-1}}$ both for the WXM (2-10 keV) and for
FREGATE (30-400 keV)
(J.L. Atteia, private communication).

Swift uses two detection modes: one based on a flux threshold and one
on fluence threshold (image mode).  The post-launch analysis can be
found in \citet{band:06}. We use his results to define a detection
probability at peak flux $P_{15-150\ \mathrm{keV}}$, based on the observed
GRB duration distribution. The mean flux threshold is $\sim 0.2\
\mathrm{ph~cm^{-2}~s^{-1}}$.  We also defined a sub-sample of ``bright
SWIFT bursts'' by selecting bursts with peak flux $P_{15-150\
\mathrm{keV}} > 1\ \mathrm{ph~cm^{-2}~s^{-1}}$.


\begin{table}
\begin{minipage}{0.5\textwidth}
\caption{\textbf{Best models: mean redshift.}}
\begin{tabular}{lccc}
\hline
\textbf{Rate} & \textbf{All} & \textbf{SWIFT} & \textbf{Bright SWIFT}\\
\hline
\multicolumn{4}{c}{\textbf{Amati-like relation $E_\mathrm{p}\propto L^{0.43}$}}\\
 & & & \\
  SFR1 & 3.1  & 1.6 & 1.3\\
  SFR2 & 8.0  & 1.9 & 1.5\\
  SFR3 & 10.5 & 3.3 & 2.1\\
\hline
\multicolumn{4}{c}{\textbf{log-normal peak energy distribution}}\\
 & & & \\
  SFR1 & 3.1  & 1.9 & 1.6\\
  SFR2 & 8.0  & 2.4 & 1.8\\
  SFR3 & 10.5 & 4.8 & 2.7\\
\hline
\end{tabular}
\end{minipage}
\label{tab:results}
\end{table}


\begin{figure*}
\begin{center}
\begin{tabular}{cc}
\includegraphics[width=0.49\textwidth]{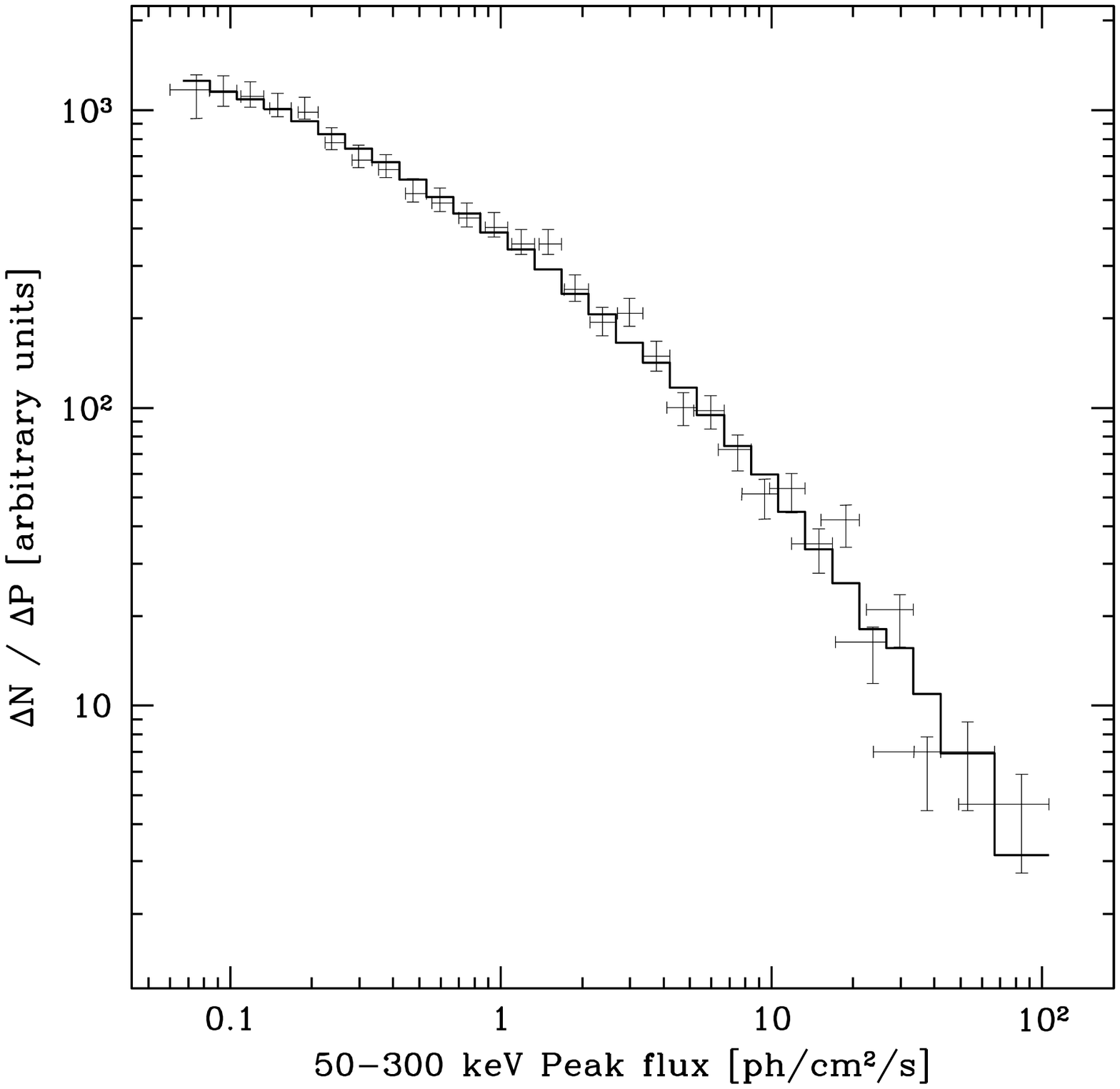} &
\includegraphics[width=0.49\textwidth]{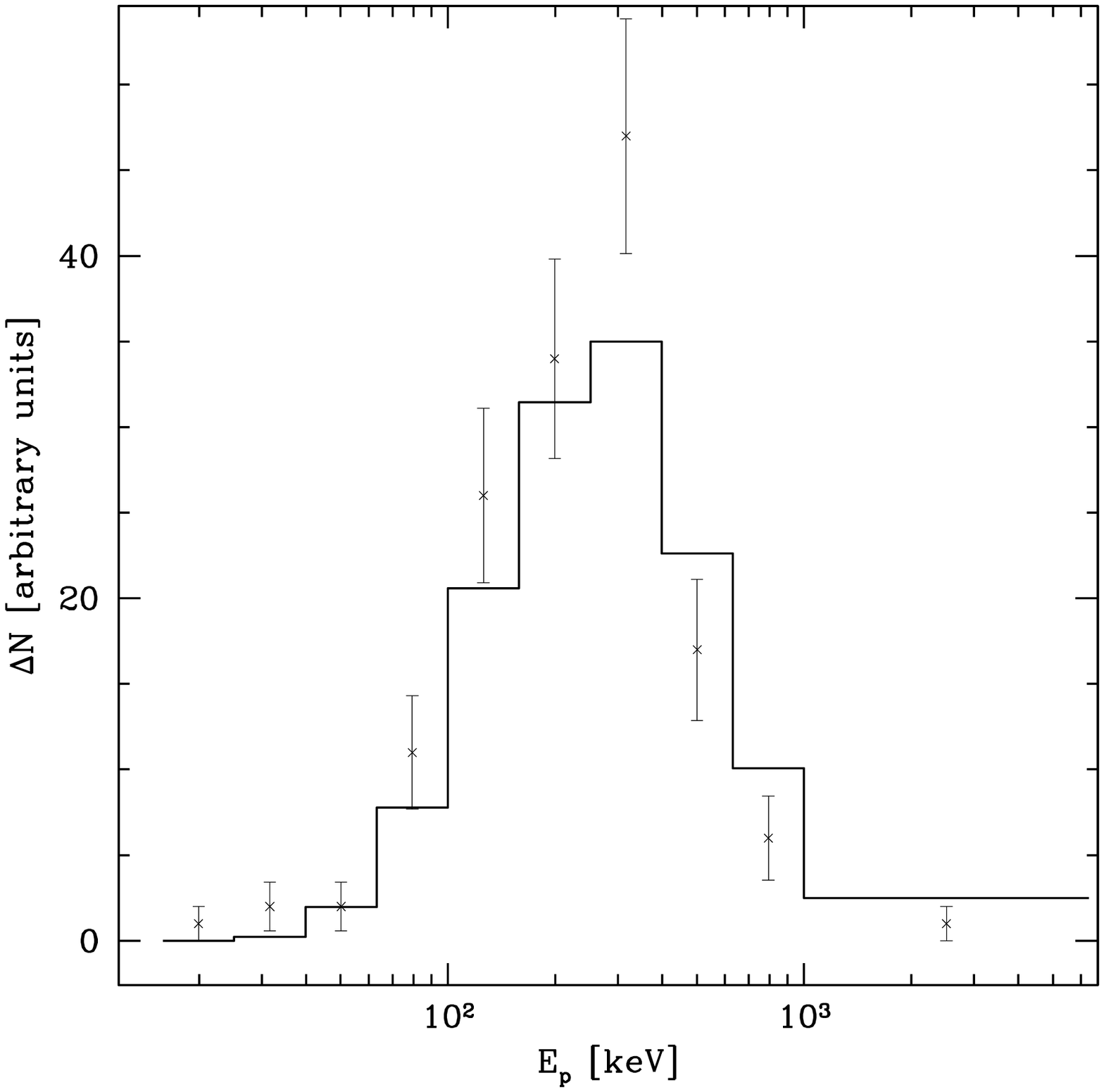}\\
\end{tabular}
\end{center}
\caption{\textbf{Best model (SFR$_{3}$, Amati-like relation)~.} \textit{Left:} the simulated $\log{N}-\log{P}$ diagram of BATSE bursts is plotted as well as BATSE data \citep{stern:00,stern:02}; \textit{right:} the simulated peak energy distribution of bright BATSE bursts is plotted as well as the observed distribution \citep{preece:00}.}
\label{fig:data}
\end{figure*}

\subsection{Observational constraints}
\label{sec:obs}

Depending on the assumption for the intrinsic peak energy
distribution, our model has four (Amati-like relation) or five
(log-normal $E_\mathrm{p}$ distribution) free parameters. These
parameters can be constrained by the following observations:
 (1) the
$\log{N}-\log{P}$ diagram of BATSE bursts
\citep{kommers:00,stern:00,stern:02}. 
(2)  the observed peak energy distribution of long
bright GRBs \citep{preece:00}
(3) the
observed fraction of XRRs and XRFs in the sample of GRBs detected by
HETE2. We adopt the same definitions as the HETE2 team for these
sub-classes, as well as their published observed ratio
\citep{sakamoto:05}.

Combining those three observations, we get 41
data points, so that our model has 37 (resp. 36) degrees of freedom in
the Amati-like relation scenario (resp. the log-normal $E_\mathrm{p}$
distribution scenario). 
We find the best fit parameters and their dispersion by $\chi^{2}$ 
minimization, as detailed below.

\subsection{Parameter determination}
\label{sec:fit}

We consider the six cases corresponding to our three assumed SFRs and
our two possible spectral scenarios. In each case, we are able to
find a good fit to the data. 
The search for the best fit is made by randomly choosing 
$L_\mathrm{min}$ (resp. $L_\mathrm{max}$) in
the interval $10^{46}$--$10^{52}\ \mathrm{erg~s^{-1}}$
(resp. $10^{51}$--$10^{56}\ \mathrm{erg~s^{-1}}$), $\delta$
in the range 0.5-3 and by adjusting the normalization $k$ to minimize
the $\chi^{2}$. More than $10^{5}$ sets of parameters are tried for
each of the six scenarios, and more than $10^{5}$ GRBs are simulated in the Monte
Carlo run for each of these sets. For each case, we define ``best models'' as models that fit the three observational constraints at the 1 sigma level, i.e. models for which $\chi^{2}_\mathrm{min} \le \chi^{2} \le \chi^{2}_\mathrm{min}+\Delta\chi^{2}$, with $\Delta\chi^{2}=40.5$ (resp. 39.5) for 37 degrees of freedom (resp. 36). We can then compute the mean value and the dispersion of the best model parameters, as well as of any function of these parameters, such as the predicted SWIFT GRB redshift distribution.
An example is given in Fig.~\ref{fig:param} in the case SFR$_{3}$ + Amati-like relation.
Very similar plots are obtained
in all the other considered cases. 
In the top-left panel of Fig.~\ref{fig:param}, we show the
location of the best models in the parameter space
$L_\mathrm{min}$--$L_\mathrm{max}$--$\delta$, using the $\log{N}-\log{P}$ constraint only. 
It clearly appears that this constraint can fix the
slope $\delta$ of the luminosity function but leaves a degeneracy on $L_\mathrm{min}$ and
$L_\mathrm{max}$. Adding the two other constraints improve the parameter determination (top-right, bottom-left panels in Fig.~\ref{fig:param}).
Indeed, it can be seen
 that 
$L_\mathrm{max}$ is better constrained using the peak energy distribution of bright BATSE bursts. 
The uncertainty on  $L_\mathrm{min}$ is slightly reduced using the third constraint.
Finally the comparison between the top-left panel ($\log{N}-\log{P}$ only) and the bottom-right panel (three constraints together) shows the improvement in the parameter determination given by our method.

In all considered cases, we always find a clear minimum for $\chi^{2}$. As expected, we also find that for each physical quantity, the mean value for the best models defined as above is very close to the value for the minimum $\chi^{2}$ model.
The quality of the best fit can be seen in Fig.~\ref{fig:data}.
Finally 
the results are shown in Table~1,
where we indicate for each case the mean value and the dispersion of the best model parameters. 
The obtained error bars confirm that the slope of the luminosity function remains better determined that the minimum and maximum luminosities.

\begin{figure*}
\begin{center}
\includegraphics[width=\textwidth]{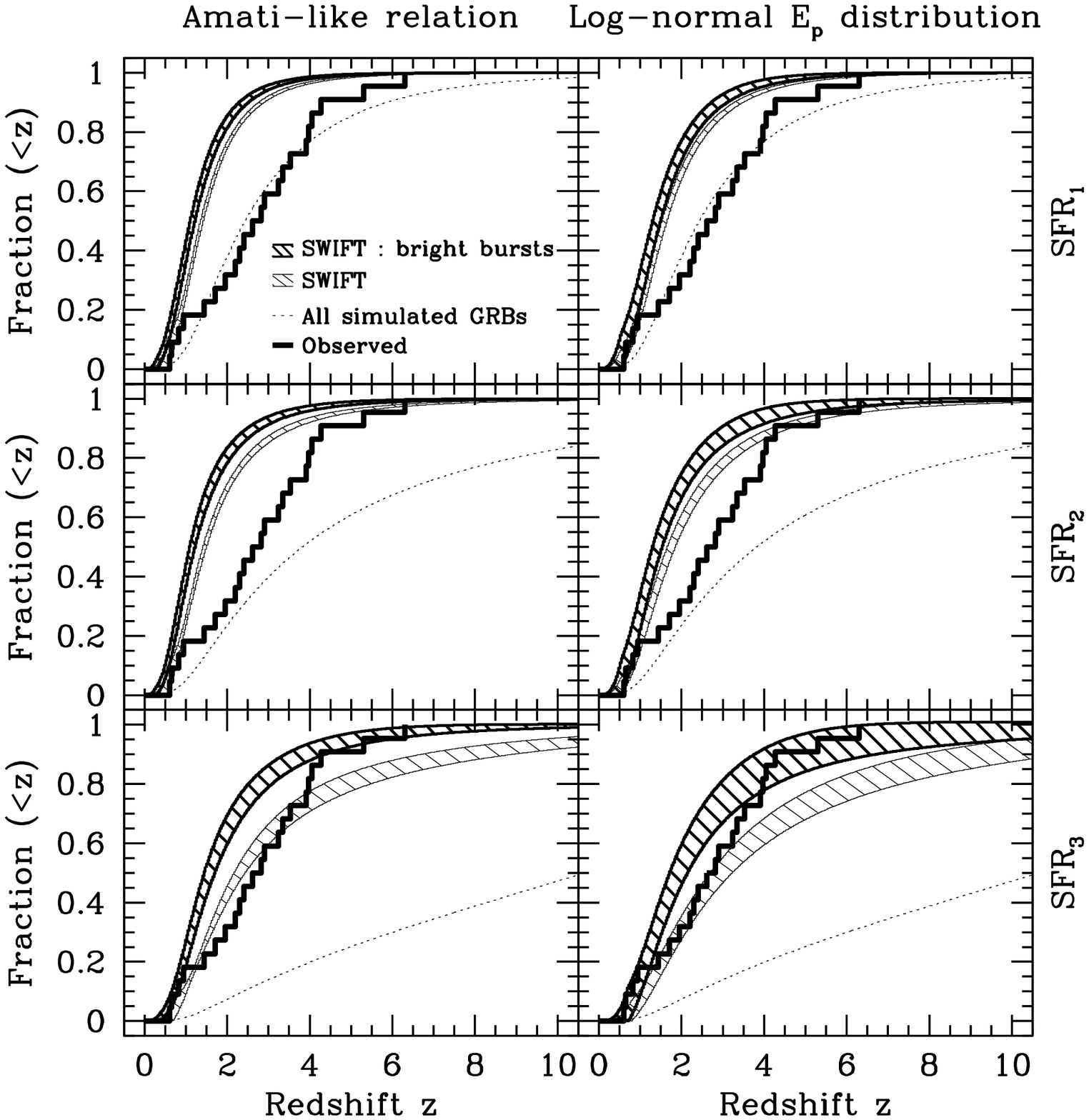}\\
\end{center}
\caption{\textbf{Best models: redshift distribution.} Each diagram
shows for the six considered cases (1) the observed distribution of
SWIFT bursts with measured redshifts (thick line); (2) the redshift
distribution of all simulated bursts (thin dotted line); (3) the
redshift distribution of simulated bursts that are detected by SWIFT
(thin hatched region); (4) the redshift distribution of bright SWIFT
bursts (thick hatched region).}
\label{fig:z}
\end{figure*}

\begin{figure*}
\begin{center}
\begin{tabular}{cc}
\textbf{Intrinsic Amati-like relation} & \textbf{Log-normal intrinsic peak energy distribution}\\
\includegraphics[width=0.49\textwidth]{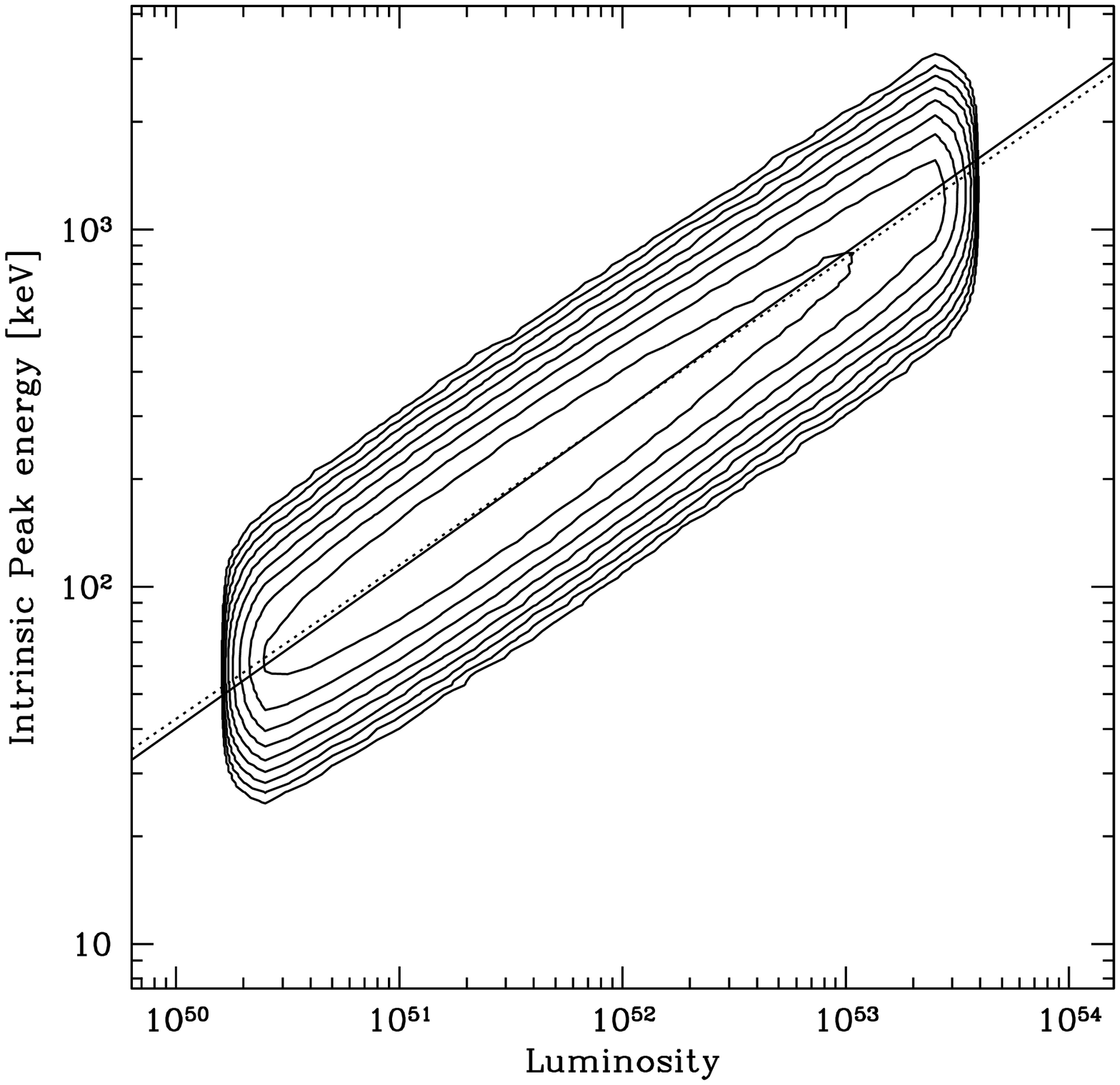} &
\includegraphics[width=0.49\textwidth]{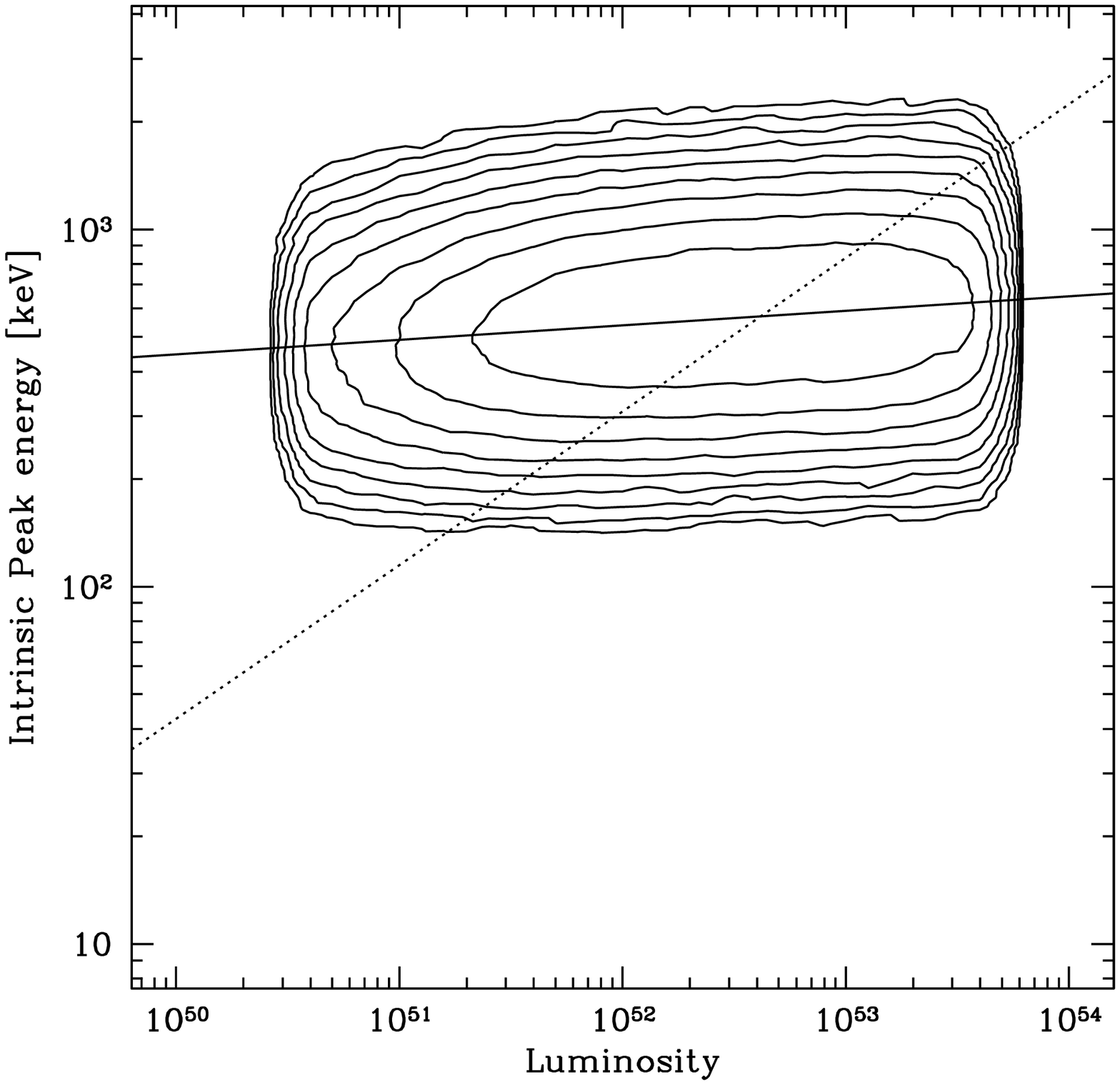}\\
\end{tabular}
\end{center}
\caption{\textbf{Amati-like relation in the best model (SFR$_{3}$).}
GRB density contours are plotted in the luminosity-intrinsic peak
energy plane for SWIFT bursts. The slope is indicated in solid line,
as well as the observed slope in dotted line
\citep{ghirlanda:05}. \textit{Left :} simulation using an intrinsic
Amati-like relation; \textit{right :} simulation where the intrinsic peak
energy follows a log-normal
distribution.}
\label{fig:amati}
\end{figure*}


\begin{table}
\begin{minipage}{0.45\textwidth}
\caption{\textbf{Best models: predicted fraction of all and bright SWIFT GRBs above $z=6$ and $z=7$.} The reported values are averaged over all ``best models''. The uncertainty in each case can be estimated from Fig.~\ref{fig:z}.}
\begin{tabular}{lcccc}
\hline
  & \multicolumn{2}{c}{\textbf{SWIFT}} & \multicolumn{2}{c}{\textbf{Bright SWIFT}}\\
\textbf{Rate} & \%($z>6$) & \%($z>7$) & \%($z>6$) & \%($z>7$)\\
\hline
\multicolumn{5}{c}{\textbf{Amati-like relation $E_\mathrm{p}\propto L^{0.43}$}}\\
 & & & \\
  SFR1 & $0.7 \%$ & $0.4 \%$ & $0.3 \%$ & $0.1 \%$\\
  SFR2 & $2.5 \%$ & $1.6 \%$ & $0.8 \%$ & $0.4 \%$\\
  SFR3 & $ 15 \%$ & $ 12 \%$ & $2.0 \%$ & $1.8 \%$\\
\hline
\multicolumn{5}{c}{\textbf{log-normal peak energy distribution}}\\
 & & & \\
  SFR1 & $1.4 \%$ & $0.7 \%$ & $0.6 \%$ & $0.3 \%$\\
  SFR2 & $4.5 \%$ & $2.9 \%$ & $1.6 \%$ & $1.0 \%$\\
  SFR3 & $ 21 \%$ & $ 17 \%$ & $6.2 \%$ & $4.5 \%$\\
\hline
\end{tabular}
\end{minipage}
\label{tab:highz}
\end{table}

\section{Results and discussion}
\label{sec:discussion}

\subsection{Redshift distribution}
\label{sec:redshift}

For each scenario, we report
in Table~2 the mean redshift we
obtain for all GRBs (intrinsic mean redshift), all GRBs detected by
SWIFT and bright GRBs detected by SWIFT. The corresponding redshift
distributions are plotted in Fig.~\ref{fig:z}, as well as the
observed distribution for SWIFT bursts \citep{jakobsson:06}. 
Each simulated redshift distribution is computed
averaging over the ``best models''.
The computed dispersion around this mean is the hatched region in Fig.~\ref{fig:z}.\\
The observed redshift distribution of SWIFT bursts is 
plagued by selection effects, difficult to account for.
For example, the deficiency of low redshift ($z\sim 1$) events, 
with respect to pre-SWIFT bursts \citep[see fig.3 in][]{jakobsson:06}
is not yet understood. 
For this reason, we do not attempt a formal fit of the data.  However,
we expect the observed distribution to be located between the simulated
distributions of all SWIFT bursts (thin hatched region in
fig.~\ref{fig:z}) and of ``bright'' SWIFT bursts, with peak flux above
$\mathrm{1\ ph~cm^{-2}~s^{1}}$ in the 15-150 keV band (thick hatched
region).\\
Figure~\ref{fig:z} clearly shows that the expected redshift
distribution strongly depends on the assumption made on the GRB
comoving rate, so that SWIFT data can already put severe constraints
on this rate.  
The case SFR$_{1}$ illustrates the effect of instrument thresholds\,: 
with a mean redshift
$\left\langle z \right\rangle \sim 1.3-1.9$,
the simulated redshift distribution of bursts detected by SWIFT is very far from the observed one 
($\left\langle z\right\rangle\sim 2.8$; \citealt{jakobsson:06}), despite an intrinsic distribution of all GRBs that is much closer.
Similarly
$\mathrm{SFR}_{2}$ gives a lower mean redshift ($\left\langle z \right\rangle \sim 1.5-2.4$, see
table~2) than the observed one.
 Conversely, $\mathrm{SFR}_{3}$ gives a much better
agreement with data ($\left\langle z \right\rangle \sim 2.1-4.8$). The
observed SWIFT distribution lies just between the simulated redshift
distribution of all SWIFT bursts and that of bright SWIFT bursts, as
expected. 
This is qualitatively in agreement with the analysis performed by \citet{jakobsson:06} who found that SWIFT data were better reproduced using Model II of \citet{natarajan:05}, which corresponds to a SFR still rising up to $z\sim 7$, i.e. close to our SFR$_{3}$.
Our results seem rather independent on the details of the
intrinsic GRB properties, as the Amati-like relation and the
log-normal $E_\mathrm{p}$ give very similar distributions.

We provide in addition in Table~3 the predicted fraction of all and bright SWIFT bursts above $z=6$ and $z=7$. We find that this fraction is less than 1 \%
for SFR$_{1}$. Assuming 100 SWIFT GRBs per year, we expect less than 5 GRBs per year above $z = 6$ for SFR$_{2}$ (resp. less than 2 bright GRBs per year). 
In this case, most high redshift GRBs will be located around $z=6-7$. In the SFR$_{3}$ case, we expect 15-20 SWIFT GRBs per year above $z=6$, including 2-6 bright GRBs. 
In this case, a non negligible fraction of GRBs is predicted above $z=7$.
Clearly, the detection of GRB 050904 at $z=6.29$ during the first year of SWIFT \citep{kawai:06} is very unlikely with SFR$_{1}$, marginally compatible with SFR$_{2}$ and more easily explained with SFR$_{3}$.\\

The fact that SFR$_{3}$ seems preferred by data is quite surprising, as this SFR
is probably unrealistic: such a high rate of star formation at large redshift
would result in an intense production of metals that would
lead to much higher metallicities in the structures and in the IGM
than observed \citep[see e.g.][]{daigne:04,daigne:05}.
Therefore, our
results provide strong evidence that the properties of GRBs or/and GRB-progenitors 
are redshift dependent. \\
One possibility is that 
the efficiency of GRB production by stars decreases with time.
The ratio $k$ of GRBs over type II supernovae would thus increase with redshift.
If the true cosmic SFR in the Universe is more similar to
SFR$_{1}$ (resp. SFR$_{2}$), then our results show  that this efficiency at $z\sim 6-7$ is $\sim
6-9$ (resp. $\sim 2-3$) higher than at $z\sim 2$ (see
Fig.~\ref{fig:sfr}). There
is indeed increasing observational evidence that the GRB rate does not
directly trace the SFR \citep{lefloch:06}.
Such an evolution could be related to many factors\,: metallicity, mass, rotation, or
binarity of the progenitors. \\
On the other hand, there could be 
 an evolution of GRB intrinsic properties with time
(e.g. more luminous GRBs at large redshift). We tested the impact of a
GRB luminosity evolution by considering a model where 
 $L_\mathrm{min}$ and $L_\mathrm{max}$ scale as
$(1+z)^{\nu}$, the slope of the luminosity function being kept
constant. We found that it is possible to reconcile SFR$_{1}$
with SWIFT data but this would imply a very strong evolution ($\nu > 2$), probably unrealistic. 
However, it is unlikely that progenitor 
characteristics would evolve without any changing in the GRB properties.
Therefore, we suspect that both evolutionary effects are, in fact, present but that our results 
are mostly due to the former.


\begin{figure*}
\begin{center}
\begin{tabular}{cc}
\textbf{Amati-like relation} & \textbf{Log-normal $E_\mathrm{p}$
distribution}\\
\includegraphics[width=0.49\textwidth]{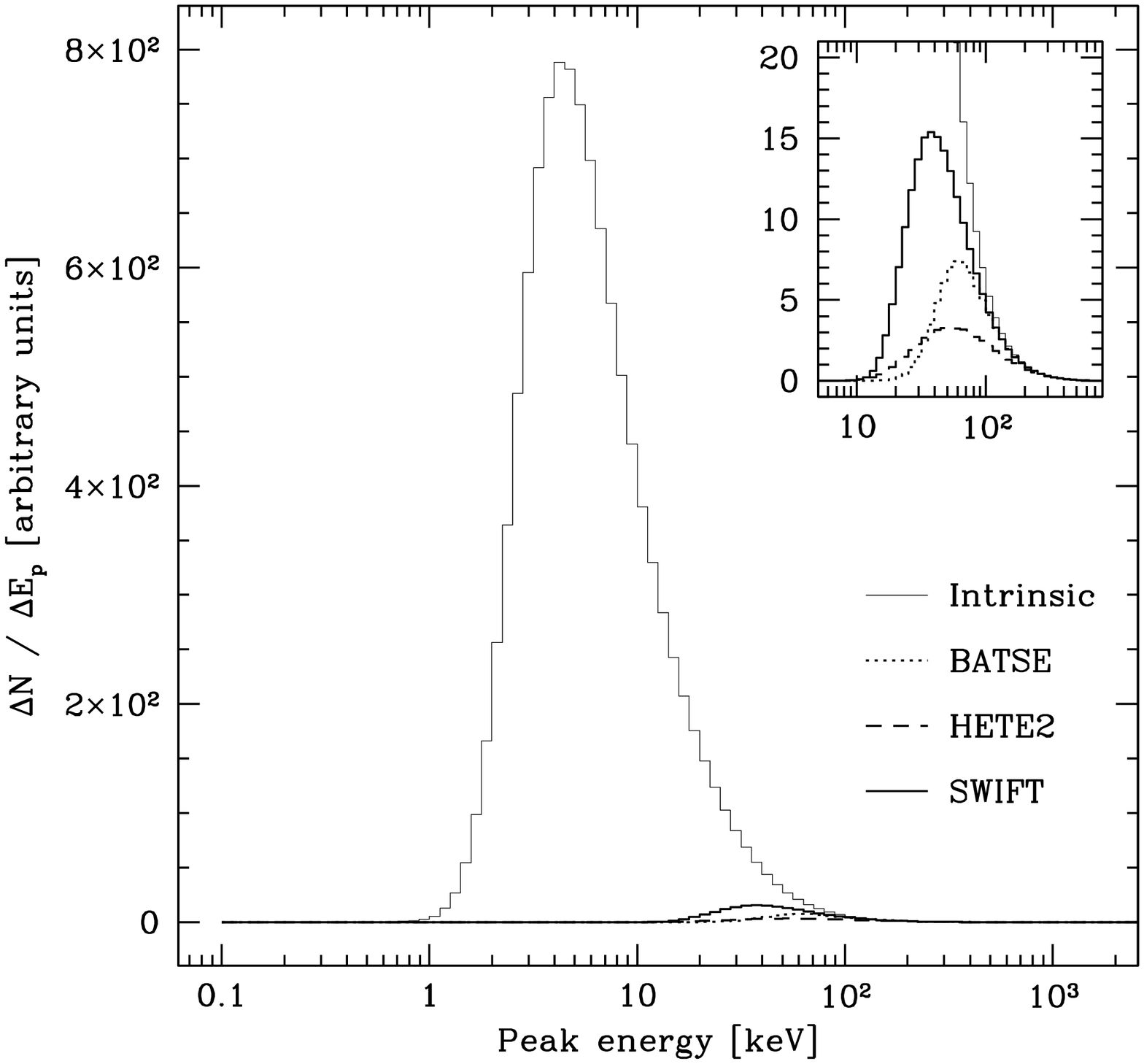} &
\includegraphics[width=0.49\textwidth]{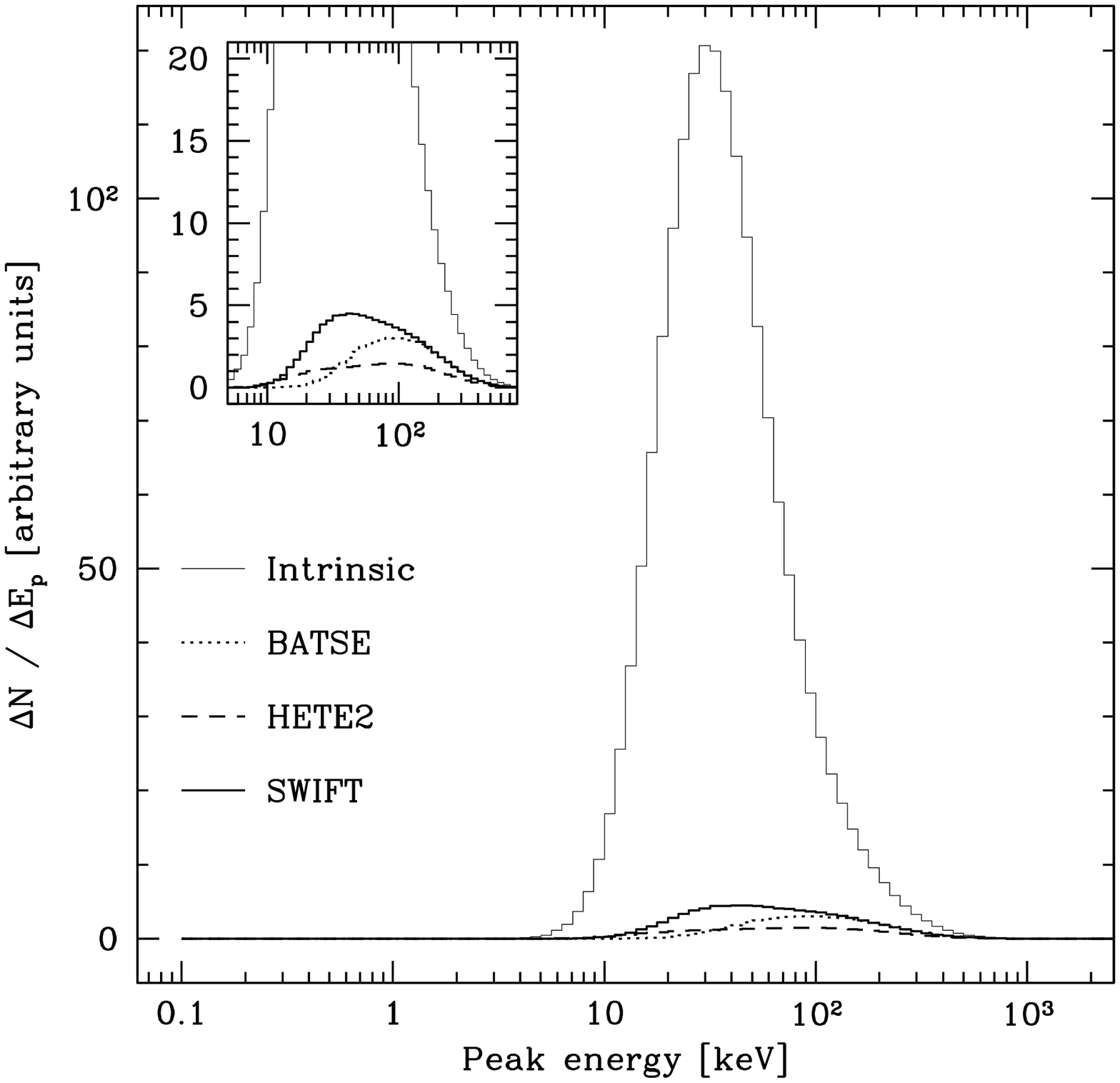}\\
\end{tabular}
\end{center}
\caption{\textbf{Observed peak energy distribution in the best model
(SFR$_{3}$).}  The peak energy distribution of all simulated GRBs, and
of simulated GRBs that are detected by BATSE, HETE2 and SWIFT are
plotted, insets showing with more details the regions of
interest. \textit{Left :} simulation using an Amati-like relation; \textit{right :}
simulation where the peak energy  follows a log-normal distribution.}
\label{fig:epeak}
\end{figure*}

\subsection{Luminosity function and peak energy distribution}

\begin{figure*}
\begin{center}
\begin{tabular}{cc}
\textbf{Amati-like relation} & \textbf{Log-normal $E_\mathrm{p}$
distribution}\\
\includegraphics[width=0.49\textwidth]{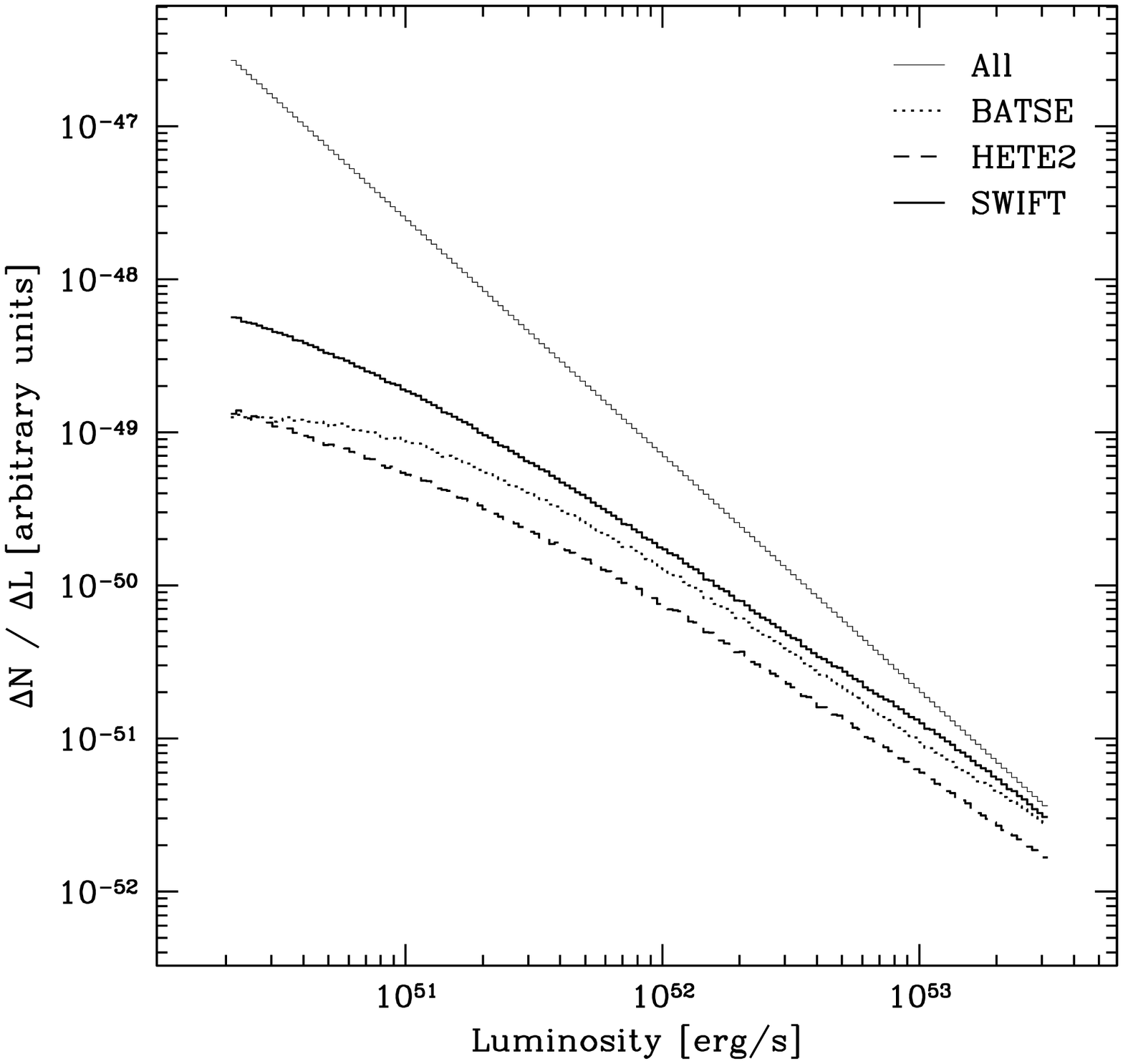} &
\includegraphics[width=0.49\textwidth]{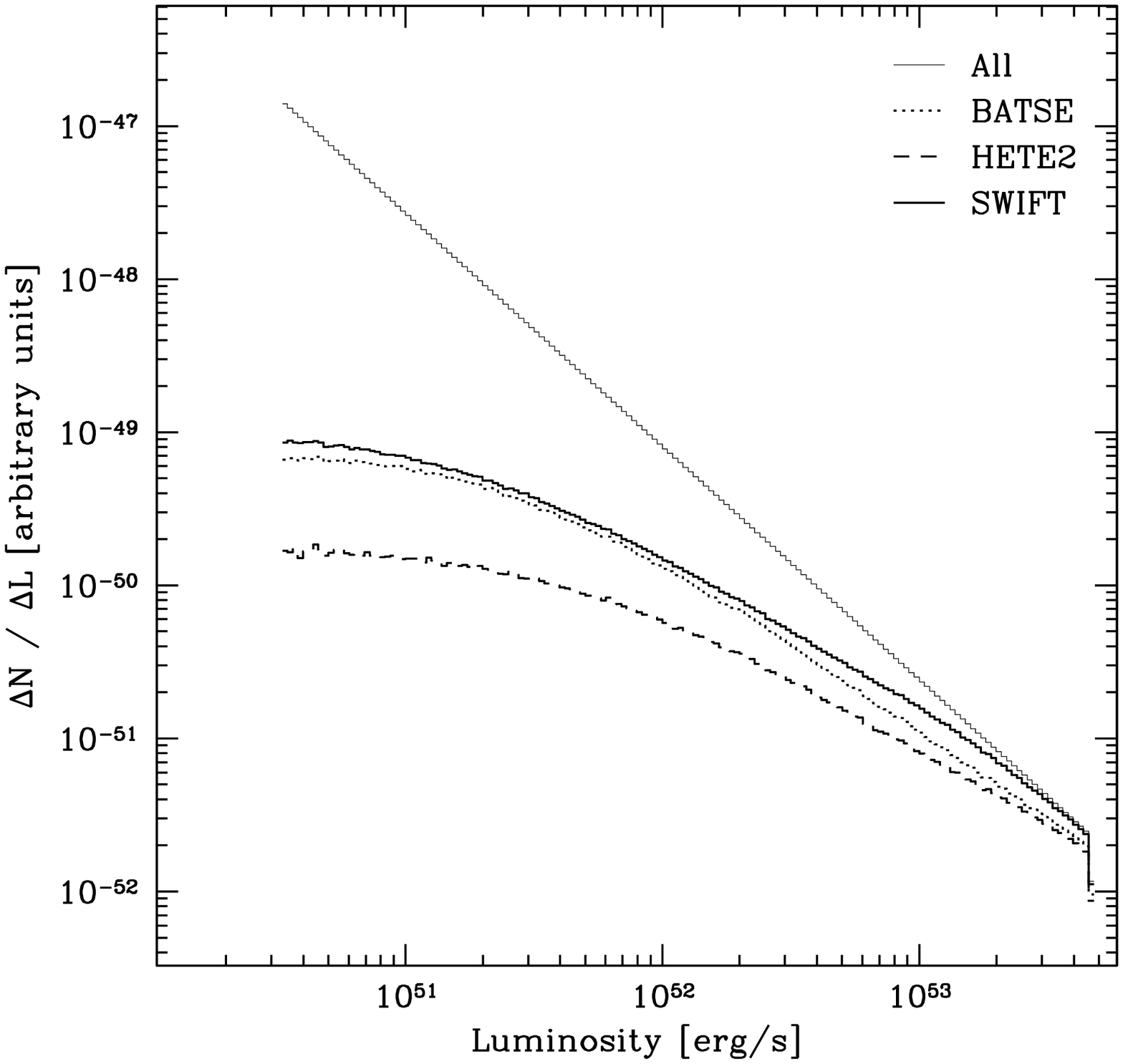}\\
\end{tabular}
\end{center}
\caption{\textbf{Luminosity function in the best model (SFR$_{3}$).}
The intrinsic luminosity function, as well as the luminosity
distribution of bursts detected by BATSE, HETE2 and SWIFT are
plotted. \textit{Left :} simulation using an intrinsic Amati-like
relation;
\textit{right :} simulation where the intrinsic peak energy  follows a log-normal distribution.}
\label{fig:luminosity}
\end{figure*}

The Amati-like relation scenario and the log-normal $E_\mathrm{p}$
distribution scenario lead to rather similar redshift distribution but
differ strongly in some other aspects: (i) as can be seen in
Fig.~\ref{fig:amati} (right panel), selection effects only can not
create an apparent Amati-like correlation between the luminosity and
the intrinsic peak energy when it is not originally present. On the
other hand, when an intrinsic correlation is assumed, detection
thresholds do not modify the observed slope (left panel). This means
that if the Amati relation is confirmed, it should have an intrinsic
origin that should be explained by GRB models; (ii)
Figure~\ref{fig:epeak} shows that the intrinsic peak energy
distribution is very different in the two scenarios. If an Amati-like
relation is present, the intrinsic peak energy distribution is
centered at about a few keV (4.5 keV for SFR$_{3}$), since 
dim bursts are more numerous. Thus, the observed peak
energy distribution in the sub-sample of bright BATSE GRBs
\citep{preece:00} is not representative of the whole GRB population, 
which is largely dominated by low $E_\mathrm{p}$  events.
On the other hand, if the peak energy
distribution is not correlated with luminosity and has a log-normal
distribution, then the observed distribution is close to the intrinsic
distribution, centered at $\sim 100\ \mathrm{keV}$. Notice that in
both cases, the mean observed peak energy by HETE2 and SWIFT is
smaller than by BATSE. This effect has also consequences on the
luminosity function. As can be seen in Fig.~\ref{fig:luminosity},
the intrinsic Amati-like relation favors the detection of low
luminosity bursts, as they have a smaller peak energy and therefore
more photons in the 15-150 keV band. In this scenario, the lower
energy threshold (15 keV vs 25 keV) of SWIFT compared to BATSE allows
SWIFT to be more efficient in sampling the luminosity function. In the
log-normal $E_\mathrm{p}$ distribution scenario, the luminosity
function of SWIFT and BATSE bursts are very similar. Notice that in both
scenarios, the observed luminosity function is far from the intrinsic
distribution, due to the small detection efficiency for the dimmest
(most numerous) GRBs. Finally note that in the log-normal distribution case, the intrinsic mean peak energy is found to be larger than the constant value assumed by PM01.\\

To check 
the dependance of our results on
 the assumed shape of the luminosity
function, we also made a few simulations using broken power-laws\,:
$$
L \propto \left\lbrace\begin{array}{cl}
L^{-0.5}    & \mathrm{for}\ L_\mathrm{min} \le L \le L_{*}\\
L^{-\delta} & \mathrm{for}\ L_{*}\le L \le L_\mathrm{max}
\end{array}\right.\ .
$$ The low-luminosity slope has been fixed to avoid an additional free
parameter.  The value $-0.5$ is a prediction of the internal shock
model \citep{daigne:06}. It happens
that the low-luminosity branch remains mostly  undetected so that the best
fit parameters are very close to those obtained with a single
power-law ($L_{*}$ adjusting itself to the previous value of
$L_\mathrm{min}$). Therefore, current data do not allow to distinguish
between single and broken power-laws for the GRB luminosity function.


\section{Conclusion}
\label{sec:conclusion}

Using Monte Carlo simulations of the long GRB population under the
assumptions that (i) the GRB rate follows the SFR; (ii) the GRB
luminosity function is a power-law; (iii) the peak-energy is either
determined by an Amati-like intrinsic relation or by a log-normal
distribution, we obtained the following results\,:

\begin{enumerate}

\item \textbf{Luminosity function:} the slope of the power-law is
well constrained, $\delta\sim 1.5-1.7$, while the minimum and maximum luminosities are not so well determined\,: $L_\mathrm{min}\sim 0.8-3\times 10^{50}\ \mathrm{erg~s^{-1}}$ and $L_\mathrm{max}\sim 3-5\times 10^{53}\ \mathrm{erg~s^{-1}}$.
The observed distribution is very biased as
most low-luminosity GRBs are not detected by current instruments.

\item \textbf{Intrinsic peak energy distribution:} if the observed
Amati relation is confirmed, the relation needs to be
intrinsic. A consequence of
the relation is that the observed peak energy distribution is not
representative of the whole GRB population, which has a lower mean
value of a few keV. Therefore, there should be many undetected XRRs and
XRFs.  On the other hand, a log-normal peak energy distribution can
also be in very good agreement with all the observational constraints,
but selection effects alone can not produce an observed Amati relation. 
In this case, the
mean peak energy of the whole GRB population is close to 100 keV and
the observed distribution is much more representative.

\item \textbf{GRB comoving rate:}
in agreement with \citet{porciani:01}, we
find that one GRB pointing towards us is produced every $ 10^{5}$
to $10^{6}$ supernovae in the Universe. 
This rate should be corrected for beaming to get the true GRB rate in the Universe. This would require 
to assume a distribution of beaming angle or to derive it using the observed correlation with the isotropic 
luminosity \citep{frail:01,ghirlanda:04}.
The present redshift distribution
of SWIFT bursts strongly favors $\mathrm{SFR}_{3}$, which still increases
above $z\sim 2$. However, SFR$_{3}$ is probably unrealistic, as so many stars at
high redshift would over-produce metals. This leads to conclude that
\textit{some GRBs or/and GRB progenitor properties evolve with redshift}.
Our a\-nalysis suggests that the main evolutionary effect could be
the redshift dependence of the efficiency of GRB production by stars, so that the GRB comoving rate still increases above $z\sim 2$, even if the SFR flattens or decreases. Thus,
{\it GRBs would not directly trace the SFR}, as also suggested by recent studies on the
properties of GRB host galaxies \citep[see e.g.][]{lefloch:06}.
To reconcile our results with a more plausible SFR such as
$\mathrm{SFR}_{1}$ or even SFR$_{2}$, this efficiency should be
about $6-7$ times larger at $z=7$ than at $z=2$. 
\item \textbf{Detection rate of high redshift GRBs:}
based on a yearly rate of $\sim 100$ SWIFT GRBs,
we predict  $\sim 2-6$ (resp. $\sim 1-5$) bright SWIFT bursts per year above $z=6$ (resp. $z=7$).
\end{enumerate}

\bibliographystyle{mn2e}
\bibliography{grbredshift}

\label{lastpage}

\end{document}